\documentclass[aps,prb,preprint,superscriptaddress]{revtex4-1}

\usepackage{graphicx}
\usepackage{braket}
\usepackage{amsmath}
\usepackage{amssymb}
\usepackage{color}
\usepackage{soul}
\usepackage[squaren,Gray]{SIunits}
\usepackage{wrapfig}
\usepackage{notes2bib}
\usepackage{float}
\usepackage{xr}
\begin{document}


\title{Self-starting harmonic frequency comb generation in a quantum cascade laser}



\author{Dmitry Kazakov$^\dagger$}

\affiliation{School of Engineering and Applied Sciences, Harvard University, Cambridge, MA 02138, USA}
\affiliation{Department of Information Technology and Electrical Engineering, ETH Zurich, 8092 Zurich, Switzerland}

\author{Marco Piccardo$^{\dagger,*}$}
\affiliation{School of Engineering and Applied Sciences, Harvard University, Cambridge, MA 02138, USA}

\author{Yongrui Wang}
\affiliation{Department of Physics and Astronomy, Texas A\&M University, College Station, TX 77843, USA}

\author{Paul Chevalier}
\affiliation{School of Engineering and Applied Sciences, Harvard University, Cambridge, MA 02138, USA}

\author{Tobias S. Mansuripur}
\affiliation{Pendar Technologies, 30 Spinelli Place, Cambridge, MA 02138, USA}

\author{Feng Xie}
\affiliation{Thorlabs Quantum Electronics (TQE), Jessup, Maryland 20794, USA}

\author{Chung-en Zah}
\affiliation{Thorlabs Quantum Electronics (TQE), Jessup, Maryland 20794, USA}

\author{Kevin Lascola}
\affiliation{Thorlabs Quantum Electronics (TQE), Jessup, Maryland 20794, USA}

\author{Alexey Belyanin}
\affiliation{Department of Physics and Astronomy, Texas A\&M University, College Station, TX 77843, USA}

\author{Federico Capasso}
\email[]{capasso@seas.harvard.edu; piccardo@g.harvard.edu}
\affiliation{School of Engineering and Applied Sciences, Harvard University, Cambridge, MA 02138, USA}


\collaboration{$^\dagger$These authors contributed equally to this work.}
\collaboration{Accepted for publication, \textit{Nature Photonics}, September 8, 2017}



\maketitle

\textbf{Optical frequency combs~\cite{Udem2002, Hansch2006} establish a rigid phase-coherent link between microwave and optical domains and are emerging as high-precision tools in an increasing number of applications~\cite{Diddams2010}. Frequency combs with large intermodal spacing are employed in the field of microwave photonics for radiofrequency arbitrary waveform synthesis~\cite{Huang2008,Wang2015} and for generation of THz tones of high spectral purity in the future wireless communication networks~\cite{Nagatsuma2016, AKYILDIZ201416}. We demonstrate for the first time self-starting harmonic frequency comb generation with a THz repetition rate in a quantum cascade laser. The large intermodal spacing caused by the suppression of tens of adjacent cavity modes originates from a parametric contribution to the gain due to temporal modulations of the population inversion in the laser~\cite{Lamb1964, Agrawal1988}. The mode spacing of the harmonic comb is shown to be uniform to within $5 \times 10^{-12}$ parts of the central frequency using multiheterodyne self-detection. This new harmonic comb state extends the range of applications of quantum cascade laser frequency combs~\cite{Hugi2012,Villares2014,Burghoff2014,Lu2015}.}

Several techniques to generate optical frequency combs (OFCs) have been demonstrated in the last decades based on different nonlinear mechanisms that fulfill the modelocking condition. Originally, passively modelocked lasers based on saturable absorption and Kerr lensing were used to create short light pulses, and were subsequently shown to also constitute frequency combs. This type of modelocking is an example of amplitude-modulated modelocking, so-named for the temporal behavior of the electric field of the emitted light. However, these techniques usually result in elaborate optical systems. More recently, new routes promising chip-scale comb generators have been investigated based on optically-pumped ultra-high-quality-factor crystalline microresonators~\cite{Kippenberg2011,Wang2013,DelHaye2016} and on broadband quantum cascade lasers (QCLs) with specially designed multistage active regions~\cite{Hugi2012,Faist2016}. In both cases the essential underlying mechanism responsible for the generation of OFCs is cascaded four-wave mixing (FWM) enabled by a third-order $\chi^{(3)}$ Kerr nonlinearity.
The temporal behavior of these OFCs is not restricted to ultrashort pulses but can represent rather sophisticated waveforms due to a non-trivial relationship among the spectral phases of the comb teeth. In fact, the output of a QCL-based frequency comb resembles that of a frequency-modulated laser with nearly constant output intensity~\cite{Hugi2012,Khurgin2014}.

\begin{figure}[H]
 \includegraphics[width=0.95\textwidth]{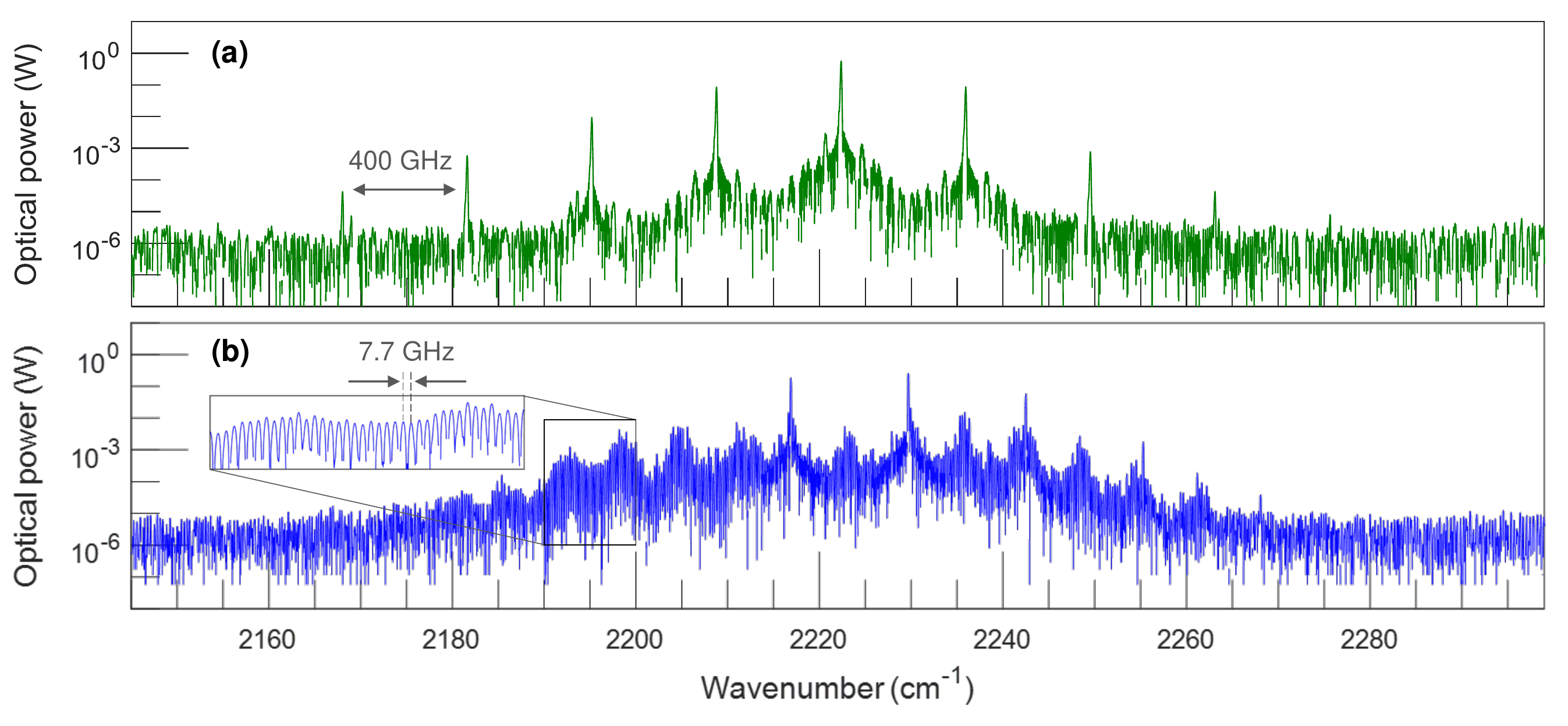}
 \caption{\textbf{Harmonically- and fundamentally-modelocked frequency combs generated in QCLs}. (a) Optical spectrum of a mid-infrared QCL in the harmonic state with a repetition rate of 400 GHz. (b) Optical spectrum of a fundamentally modelocked QCL with a repetition rate of 7.7 GHz. In both cases the cavity free spectral range is 7.7 GHz.}
 \label{spectra}
 \end{figure}

A novel mechanism of OFC generation in QCLs was suggested by the recent discovery of a new laser state~\cite{Mansuripur2016}, which comprises many modes separated by higher harmonics of the cavity free spectral range (FSR) (Figure~\ref{spectra}a). This spectrum radically differs from that of fundamentally modelocked QCL combs where adjacent cavity modes are populated (Figure \ref{spectra}b). This new state is achieved by controlling the current so that the QCL first reaches a state of high single-mode intracavity intensity. When this intensity is large enough, an instability threshold is reached caused by the $\chi^{(3)}$ population pulsation nonlinearity, favoring the appearance of modes separated by tens of FSRs from the first lasing mode. In this work, we verify the equidistance of these modes, thereby proving that QCLs are capable of harmonic modelocking and concomitant high-repetition-rate OFC generation. While OFCs with repetition frequencies in the range between 10 and 1000 GHz have already been demonstrated in optically-pumped microresonators~\cite{Kippenberg2011}, the generation of high-repetition-rate OFCs based on harmonically modelocked QCLs presents the advantage of a truly monolithic, electrically-driven source. It is notable that this type of modelocking does not require additional intracavity nonlinear elements such as saturable absorbers, or mode-selection elements, such as Bragg reflectors, that were used to achieve passive harmonic modelocking at THz repetition rates in other semiconductor lasers~\cite{Arahira1996}. Rather, the modes are locked passively due to the behavior of the QCL gain medium itself.

In this work we employed two Fabry-Perot (FP) QCLs fabricated from the same growth process with 6 mm-long cavities and emitting at 4.5 $\mu$m (see Methods for details). The devices are coated with a high-reflectivity coating ($\mathrm{R}\approx1$) on the back facet and an anti-reflection coating ($\mathrm{R}\approx0.01$) on the front facet.  Due to the cavity asymmetry introduced by the coatings, the harmonic spectra produced by these lasers exhibit sidebands with much larger separation than that of nominally identical uncoated devices, a phenomenon attributed to a weaker population grating favoring a coherent instability with larger sideband separation~\cite{Mansuripur2016}.

\begin{figure}[H]
\centering
\includegraphics[width=0.5\textwidth]{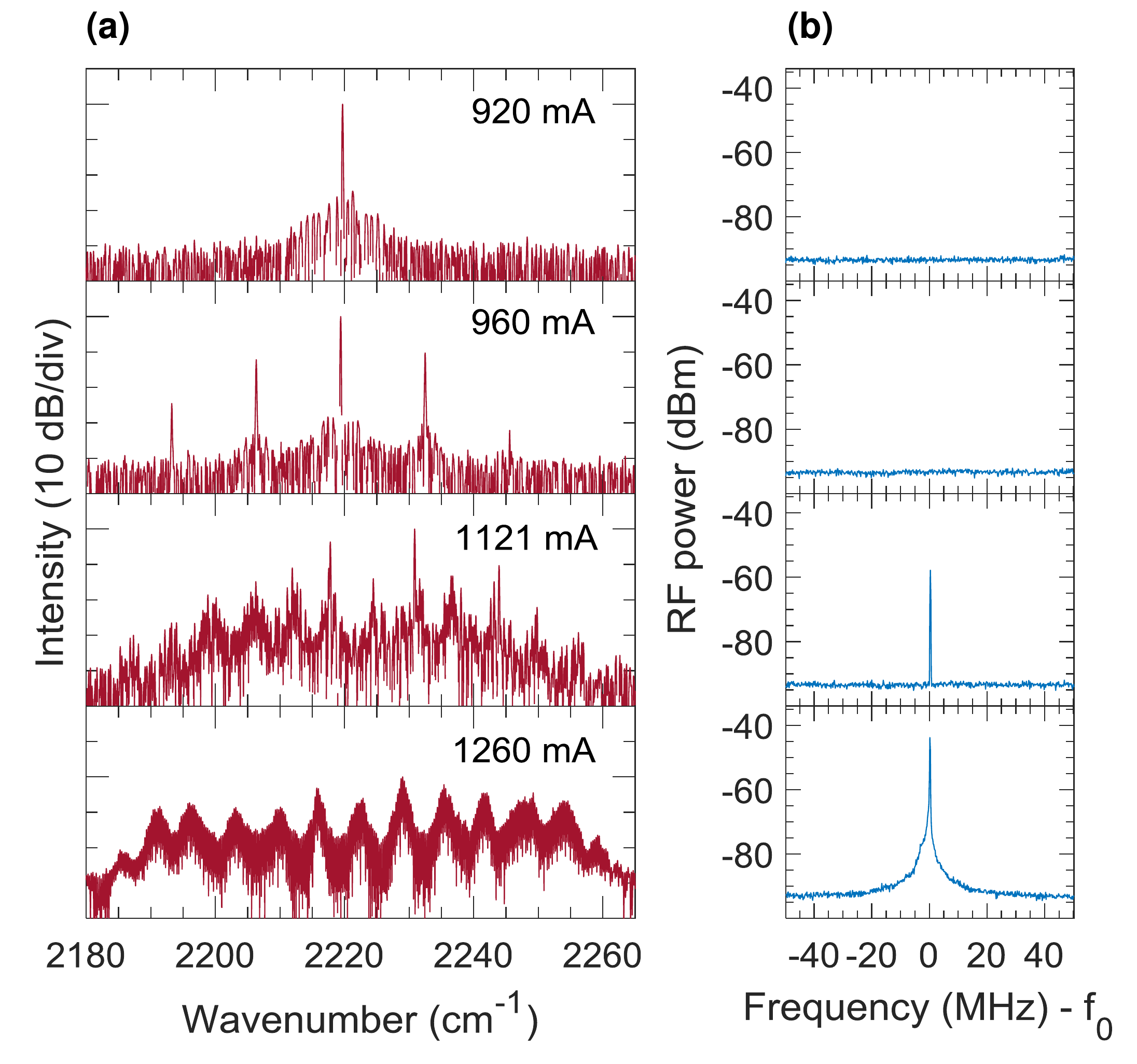}
\caption{\textbf{Spectral evolution of Fabry-Perot QCLs}. (a) Optical spectra corresponding to different laser states (from low to high current): single mode, harmonic frequency comb, fundamentally modelocked frequency comb, and high-phase-noise state. (b) Corresponding radiofrequency spectra acquired at the QCL cavity roundtrip frequency ($f_0=7.6776$ GHz at 920, 960 and 1121 mA, and $7.6678$ GHz at 1260 mA, $\mathrm{RBW}=200$ kHz). The absence of the intermodal beatnote in the harmonic state (960 mA) signifies the suppression of adjacent FP modes. The narrow beatnote at the cavity FSR is a signature of comb operation and appears when neighboring FP modes start lasing (1121 mA). At 1260 mA the comb is destabilized and the beatnote features a high-phase-noise pedestal.}
\label{regimes}
\end{figure}

These devices exhibit four distinct laser states as a function of the injected current which are shown in Figure \ref{regimes}a.   
Starting from single-mode operation and slowly increasing the bias, one can observe the harmonic state appearing at a pump current only fractionally higher than the lasing threshold. No beatnote at the cavity roundtrip frequency ($f_\text{rt}$) is observed in this regime (Figure \ref{regimes}b), confirming the absence of interleaving FP modes.
At higher values of injected current the laser transitions to a single-FSR-spaced state producing a single narrow intermodal beatnote (FWHM $< 1$ kHz) at $f_\text{rt}$ indicating the occurrence of fundamental modelocking and the comb nature of this state~\cite{Hugi2012}. Finally at even higher current the single-FSR comb acquires a high-phase-noise pedestal -- a typical signature of comb destabilization~\cite{Villares2016}.   
To verify the spacing uniformity of the modes of the harmonic state, techniques developed to investigate combs with an intermodal spacing in the lower GHz range, such as intermode beat spectroscopy \cite{Hugi2012} and SWIFTS \cite{Burghoff2014}, cannot be applied because the THz-scale beatnote frequency of the harmonic state is beyond the bandwidth of conventional mid-IR detectors and radiofrequency (RF) electronics. Instead, we use a technique that was first developed to characterize high-repetition-rate microresonator combs, in which the  sample comb spectrum is downcoverted from the optical to the RF domain by means of multiheterodyne beating with the modes of a finely-spaced reference comb~\cite{DelHaye2007}. In this scheme, if the harmonic state constitutes a frequency comb the down-converted spectrum will form an RF comb whose equidistant spacing can be accurately verified using electronic frequency counters.

\begin{figure}[H]
\includegraphics[width=\textwidth]{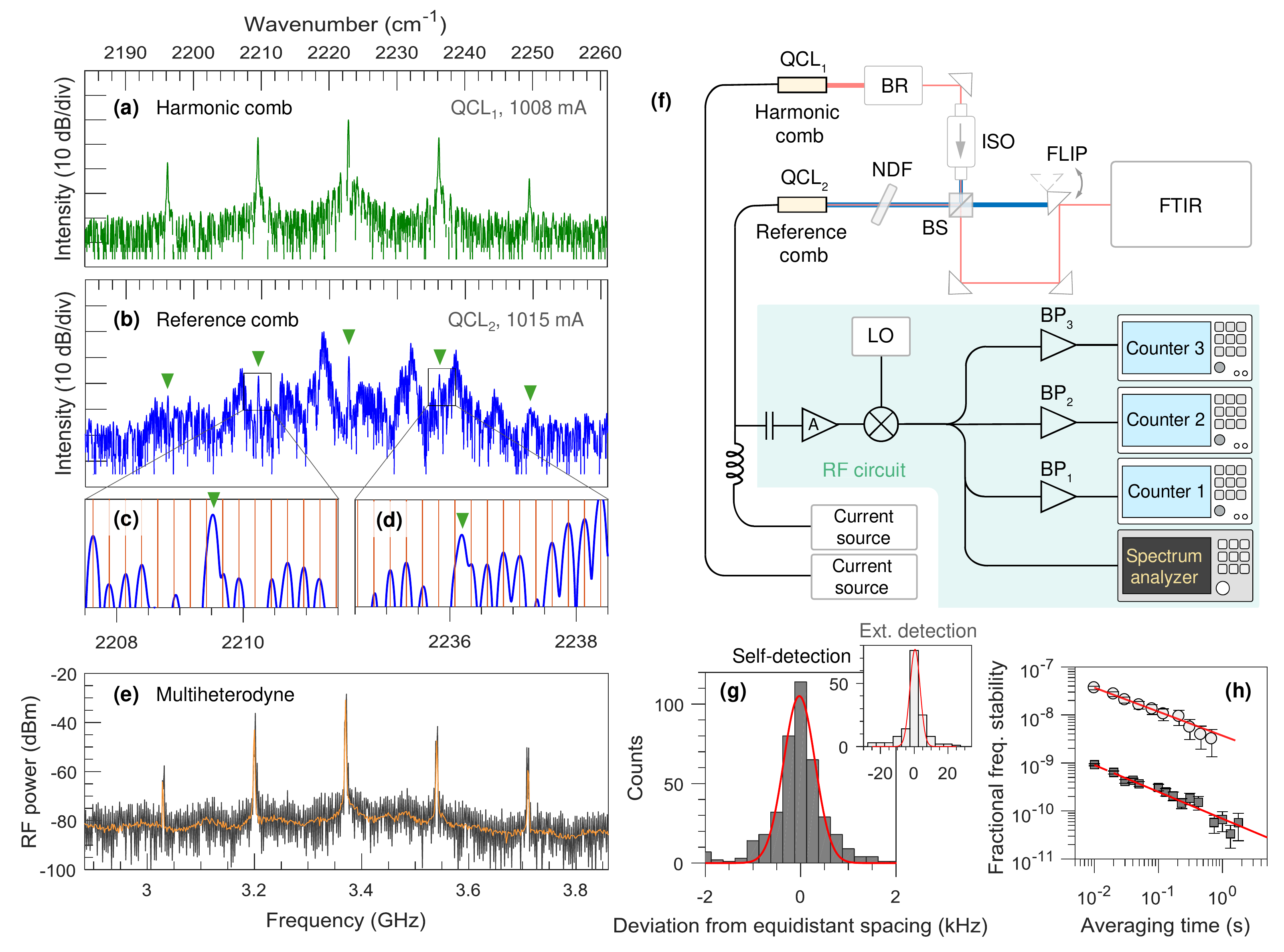}
\caption{\textbf{Mode spacing uniformity of the harmonic state}. (a) Optical spectrum of QCL$_1$ in the harmonic regime. (b) Optical spectrum of QCL$_2$ (reference comb) upon optical injection from QCL$_1$. (c) and (d) show a magnified view of a portion of the spectrum of QCL$_2$. (e) Multiheterodyne spectrum produced by the pairwise beating of the nearest modes of the two lasers. The orange trace represents the spectrum averaged over 1000 sweeps of 2 ms. (f) Optical and RF set-up for the assessment of the comb nature of the harmonic state. BR, beam reducer; ISO, optical isolator; BS, beamsplitter; FLIP, flip mirror; NDF, neutral density filter; BP, bandpass filter. The RF mixer is used to downconvert the multiheterodyne signal and aligh the beatnotes within the passbands of the electrical filters by changing the LO frequency. (g) Histogram showing the deviation from equidistant spacing of the harmonic state measured with the set-up shown in (f) for a gate time of 10 ms and 402 counts. The parameters of the Gaussian fit (red curve) are $\mu=-27$ Hz and $\sigma=329$ Hz. Inset: histogram obtained with the same technique while beating on external detector. (h) Fractional frequency stability of the dual-comb system in self-detection mode (squares) versus external detection mode (circles). The self-detected system shows more than an order of magnitude of increase in frequency stability.}
\label{heterodyne_injection}
\end{figure}

For the multiheterodyne experiment we use two QCLs, one operating in the harmonic regime (QCL$_1$) and the other in a fundamentally modelocked regime (QCL$_2$) acting as a reference comb with an intermodal spacing of $\sim$7.7 GHz. We employ a self-detection scheme in which the light emitted from QCL$_1$ is injected, after passing through an optical isolator, into the cavity of QCL$_2$ (Figure~\ref{heterodyne_injection}f). The latter acts at the same time as a reference comb and a fast photomixer~\cite{Rosch2016}, from which we can extract electrically the multiheterodyne signal generated by the intracavity beating of the optical fields of the two lasers. The attractive feature of this scheme is that it provides better signal stability as compared to a standard approach utilizing an external fast photodiode. The description of this method is relegated to the Supplementary Materials while its result is given for comparison in Figure~\ref{heterodyne_injection}g,h. 

The emission spectra of the harmonic and reference comb measured using the self-detection scheme are shown in Figure~\ref{heterodyne_injection}a,b. QCL$_1$ is operating in the harmonic regime, while QCL$_2$ is operating in a fundamentally modelocked regime exhibiting adjacent cavity modes that constitute an equidistant grid with a spacing defined by $f_{rt}$ (Figure~\ref{heterodyne_injection}c,d). Interestingly, several prominent peaks not lying on this grid can be identified in the spectrum of QCL$_2$ (marked by green triangles in Figure~\ref{heterodyne_injection}b-d) corresponding to modes injected from QCL$_1$ into QCL$_2$.
The pairwise beating of the modes of the harmonic state with the nearest modes of the reference comb produces the multiheterodyne spectrum shown in Figure~\ref{heterodyne_injection}e.
 
To assess the locking of the harmonic modes we further downconvert the multiheterodyne signal with an RF mixer and select three beatnotes of the spectrum using bandpass filters whose output is fed into three synchronized frequency counters. From the measured frequencies a histogram showing the statistics of the deviation from equidistant spacing of the RF comb can be constructed (Figure~\ref{heterodyne_injection}g). The fractional frequency stability of the dual-comb system exhibits an inverse square root dependence on the averaging time indicating the dominance of white-noise frequency modulations in the system giving origin to random and uncorrelated fluctuations following a normal distribution (Figure~\ref{heterodyne_injection}h). This allows to fit the histogram with a Gaussian function which yields a mean value of $\mu =-27$ Hz and a standard deviation of $\sigma=329$ Hz. This result verifies the equidistant spacing of the harmonic comb with a relative accuracy of $\sigma/f_{\text{c}}=5\times10^{-12}$, as normalized to the optical carrier frequency of the laser ($f_{\text{c}} = 66.7$ THz), being an order of magnitude smaller than for the measurement based on external detection. This net improvement is due to the higher stability of the multiheterodyne signal in the self-detection scheme (Figure~\ref{heterodyne_injection}h).

\begin{figure}[H]
\centering
\includegraphics[width=0.5\columnwidth]{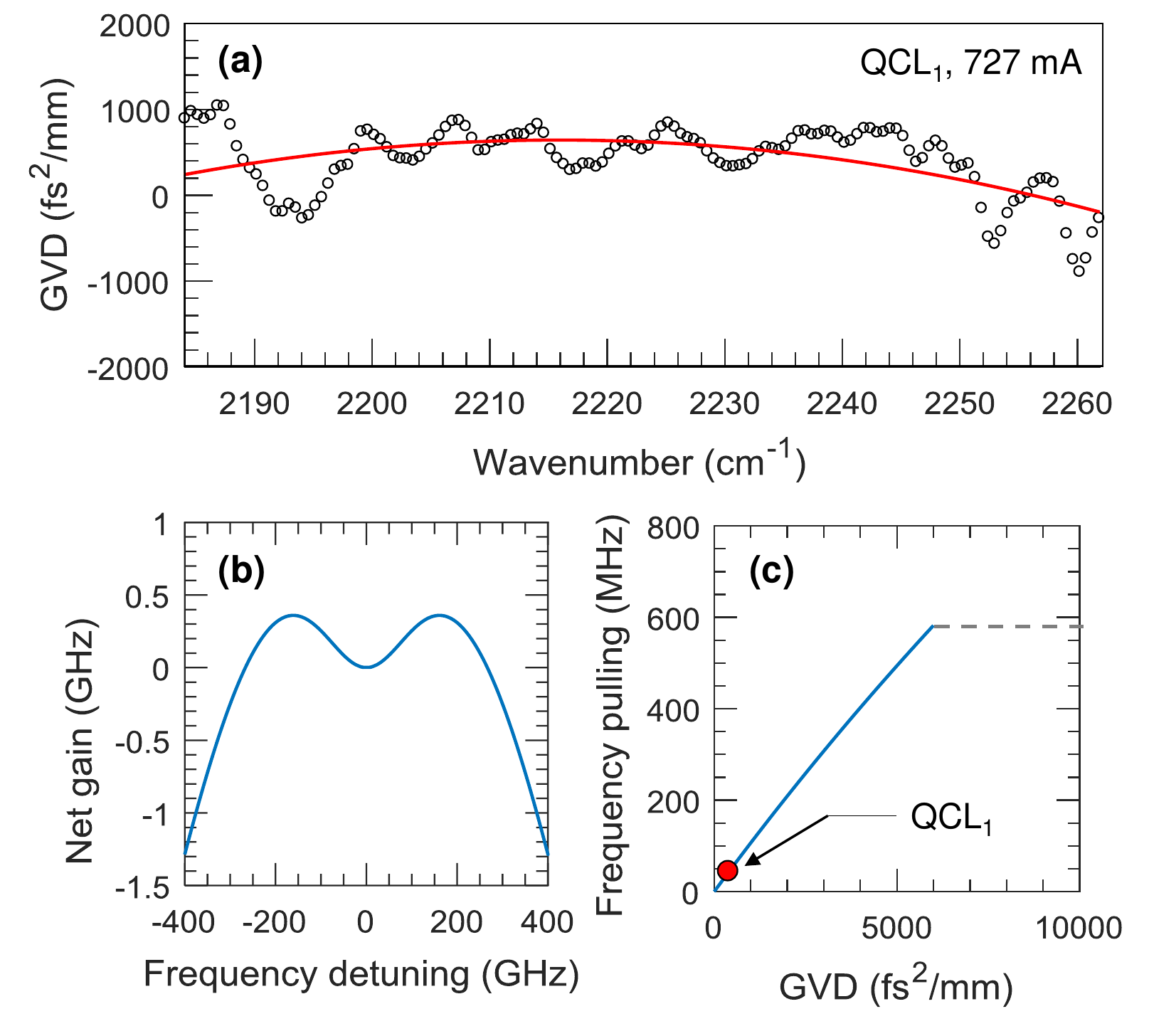}
\caption{\textbf{Harmonic modelocking in presence of native dispersion}. (a) Measured subthreshold GVD of $\mathrm{QCL}_1$. Experimental data (empty circles) is fitted with a parabola (red curve). (b) Net gain predicted by the perturbation theory of harmonic comb formation using the device parameters of QCL$_1$, including the measured subthreshold GVD. (c) Calculated frequency pulling exerted by FWM processes occurring in the QCL and compensating the dispersion of the laser as a function of the subthreshold GVD (continuous blue line), a phenomenon theoretically treated in Supplementary Section~III. Above 6000 fs$^2$/mm FWM cannot compensate for the laser dispersion (dashed grey line) preventing the occurrence of harmonic modelocking. The GVD value of QCL$_1$ is shown as a reference (red circle).}
\label{theory}
\end{figure}

In order to explain the occurrence of harmonic modelocking we resort to a perturbation theory of comb formation in QCLs considering the interaction of a two-level gain medium with a field comprised by a central mode and two weak equally detuned sidebands in the laser cavity. The nature of the parametric gain responsible for adjacent mode skipping in the laser was already studied in Ref.~\citenum{Mansuripur2016}. Here we  apply a more general approach which includes in a  systematic way the effects of nonequal sideband amplitudes, diffusion of the population grating and group velocity dispersion (GVD), which may hamper comb operation~\cite{Villares2016}. The complete derivation of our theory is given in the Supplementary Section~III, while here we outline the main implications given by the solutions of our model for the real device parameters. The subthreshold GVD of QCL$_{1}$ measured by a standard technique~\cite{Hofstetter1999} is displayed in Figure~\ref{theory}a. The net parametric gain calculated as a function of sideband detuning is shown in Figure~\ref{theory}b: it peaks at a frequency of 200 GHz (26 FSR of a 6 mm long cavity) with respect to the central mode indicating that the modes at this frequency are the first to oscillate, while the modes lying closer to the central pump are parametrically suppressed. Furthermore we calculate that FWM can compensate for the non-zero dispersion of the real device up to a value of GVD of 6000 fs$^2$/mm above which the onset of harmonic modelocking is hampered (Figure~\ref{theory}c). These results are consistent with our experimental findings proving the occurrence of harmonic modelocking in a QCL with GVD below 1000 fs$^2$/mm.

The ability to generate and passively mode-lock harmonic modes  while skipping adjacent cavity resonances relies on a coherent instability enabled by the QCL gain medium itself unveiling the compelling dynamics of the new laser state in QCLs. Locking of comb teeth with a spacing comparable to the gain recovery frequency represents a major step towards the demonstration of coherent mid-infrared amplitude-modulated waveform emission from QCLs, long thought to be prevented by the underlying physical principles, paving the way towards applications requiring short pulses of mid-infrared light. On the other hand, QCL harmonic comb generators may find their application in future wireless THz communication networks~\cite{Petrov2016}, as they combine the functionality of a high-bandwidth photomixer and a comb source promising the intracavity generation of powerful THz carrier signals, whose frequency can be designed by engineering the facet coatings of the device~\cite{Mansuripur2016} and where the phase noise is inherently low due to a high degree of correlation among the optical modes that produce the beatnote. Merging this capability with the fact that QCLs can be optimized to have a flat frequency response over a large modulation bandwidth~\cite{Hinkov2016} may allow them to operate as compact unibody modems to transmit and receive digital data in the THz communication band.

\section*{Methods}
\textbf{Quantum cascade lasers}. The devices are continuous wave, buried heterostructure, Fabry-Perot QCLs fabricated from the same growth process and emitting at 4.5 $\mu$m (Thorlabs). The single-stack active region consists of strain-balanced Ga$_\mathrm{x}$In$_{1-\mathrm{x}}$As/Al$_\mathrm{y}$In$_{1-\mathrm{y}}$As layers grown on an InP substrate~\cite{Xie2011}. The length and width of the buried waveguide are, respectively, 6 mm and 5 $\mu\mathrm{m}$. The spectral evolution of the device named QCL$_1$ was reported in Ref.~\citenum{Mansuripur2016} (``TL-4.6:HR/AR") while that of QCL$_2$ is given in the Supplementary Section~I. The QCLs are driven with low-noise current drivers (Wavelength Electronics QCL LAB 2000) with an average specified current noise density of 4 nA/$\mathrm{\sqrt{Hz}}$ and their temperature is stabilized using low-thermal-drift temperature controllers (Wavelength Electronics TC5) with typical fluctuations smaller than 10 mK.

\textbf{Multiheterodyne set-up}. We employ two different configurations for the multiheterodyne experiment: the self-detection and the external detection mode. In the first scheme (Figure~\ref{heterodyne_injection}f) the beam emitted from a QCL operating in the harmonic regime (I$_{\mathrm{QCL}_1}=1008$ mA) is collimated using an off-axis parabolic mirror (25.4 mm focal length) and sent through a Faraday isolator (Innovation Photonics, 30 dB extinction ratio) to prevent feedback-induced destabilization of the harmonic state. A beam reducer is used to decrease the beam diameter and maximize its transmission through the isolator. After partial attenuation by a neutral density filter the beam is focused inside the cavity of a second QCL acting simultaneously as a reference comb and a fast photomixer (I$_{\mathrm{QCL}_2}=1015$ mA). By using a beamsplitter (45:55 splitting ratio) and a flip mirror one can selectively measure with an FTIR spectrometer (Bruker Vertex 80v, 0.1 cm$^{-1}$ resolution) the optical spectra of the harmonic comb, and reference comb upon injection. In the external detection mode (Figure S~3a) both QCLs are free-running, one operating in the harmonic regime (I$_{\mathrm{QCL}_1}=1139$ mA) and the other as a reference comb (I$_{\mathrm{QCL}_2}=1084$ mA), and the two collinear beams are focused onto an external fast MCT detector (Vigo PVI-2TE-5, 1 GHz bandwidth). The detector is tilted at an angle to minimize optical feedback on the lasers. In both configurations, the beating of the optical fields, whether occurring inside the cavity of the QCL detector or on the fast photodetector, is converted into an RF signal that is sent to the RF circuit shown in Figure~\ref{heterodyne_injection}f. An RF bias-tee is used to extract the multiheterodyne signal from the QCL detector. The RF signal is amplified and then down-converted using an RF mixer. The LO signal is supplied by a tunable signal generator (R\&S SMF100A). While monitoring the RF signal on the spectrum analyzer it is possible to tune the central frequency of the multiheterodyne spectrum by adjusting the LO frequency and the spacing between individual tones by tuning the currents of the two QCLs. This allows to align three adjacent beatnotes of the spectrum within the passbands (60 MHz) of three homemade filters, centered  at 50 MHz, 135 MHz and 220 MHz. The filtered signals are amplified and their frequency is measured using three synchronized frequency counters (Agilent 53220A) sharing the same external gate and trigger control and a 10 MHz clock reference. 
 
\textbf{Acknowledgments}

This work was supported by the DARPA SCOUT program through Grant no. W31P4Q-16-1-0002. We acknowledge support from the National Science Foundation under Award No. ECCS-1614631. Any opinions, findings, conclusions or recommendations expressed in this material are those of the authors and do not necessarily reflect the views of the Assistant Secretary of Defense for Research and Engineering or of the National Science Foundation. M.P. and D.K. wish to thank J.B. MacArthur for the assembly of RF filters, A.Y. Zhu for sputtering gold on a QCL submount and N. Rubin for a careful reading of this manuscript.

\textbf{Author contributions}

D.K. and M.P. conceived, designed and implemented the experiments and wrote the manuscript with feedback from the other co-authors. P.C. contributed to the realization of the experiments. F.X., C.Z. and K.L. provided the quantum cascade laser devices. Y.W. and A.B. developed the theoretical model and contributed to the theoretical part of the manuscript. T.S.M., P.C., A.B., M.P., D.K. and F.C. discussed the data. All work was done under the supervision of F.C. 

\textbf{Additional information}

The authors declare no competing financial interests. Correspondence should be addressed to M.P. and F.C.

\newpage

\section*{Figures}

 \newpage 
 
\bibliography{harmcombbib}

\end{document}



\title{Supplemental Material to: Self-starting harmonic frequency comb generation in a quantum cascade laser}



\author{Dmitry Kazakov$^\dagger$}

\affiliation{School of Engineering and Applied Sciences, Harvard University, Cambridge, MA 02138, USA}
\affiliation{Department of Information Technology and Electrical Engineering, ETH Zurich, 8092 Zurich, Switzerland}

\author{Marco Piccardo$^{\dagger,*}$}
\affiliation{School of Engineering and Applied Sciences, Harvard University, Cambridge, MA 02138, USA}

\author{Yongrui Wang}
\affiliation{Department of Physics and Astronomy, Texas A\&M University, College Station, TX 77843, USA}

\author{Paul Chevalier}
\affiliation{School of Engineering and Applied Sciences, Harvard University, Cambridge, MA 02138, USA}

\author{Tobias S. Mansuripur}
\affiliation{Pendar Technologies, 30 Spinelli Place, Cambridge, MA 02138, USA}

\author{Feng Xie}
\affiliation{Thorlabs Quantum Electronics (TQE), Jessup, Maryland 20794, USA}

\author{Chung-en Zah}
\affiliation{Thorlabs Quantum Electronics (TQE), Jessup, Maryland 20794, USA}

\author{Kevin Lascola}
\affiliation{Thorlabs Quantum Electronics (TQE), Jessup, Maryland 20794, USA}

\author{Alexey Belyanin}
\affiliation{Department of Physics and Astronomy, Texas A\&M University, College Station, TX 77843, USA}

\author{Federico Capasso}
\email[]{capasso@seas.harvard.edu; piccardo@g.harvard.edu}
\affiliation{School of Engineering and Applied Sciences, Harvard University, Cambridge, MA 02138, USA}


\collaboration{$^\dagger$These authors contributed equally to this work.}
\collaboration{Accepted for publication, \textit{Nature Photonics}, September 8, 2017}



\maketitle

\section{Detailed spectral evolution of the QCL}
\label{sec-QCL2evolution}

The purpose of this section is to provide further details on the spectral evolution of the QCL from single mode to dense state as a complement to the main text where we highlight the fundamental different regimes exhibited by the laser. During characterization the device (QCL$_2$) is free-running and operating at a temperature of 16$\degree$C. One can clearly identify four distinct laser states as a function of the injected current: single mode (840 mA and 920 mA), harmonic state (861-910 mA and 921-960 mA), single-FSR-spaced comb state (966-1220 mA) and high-phase-noise state (above 1260 mA). Once the instability threshold is reached the central laser mode develops two sidebands (861 mA) that proliferate further via FMW preserving the initial sideband spacing ($\sim$25 FSR) at higher currents. A remarkable event happens at 920 mA when the laser collapses back to the single mode regime (note the detuning of the mode by 10 cm$^{-1}$ with respect to the frequency at the lasing threshold) and soon after, at 921 mA, it reaches a second instability threshold with sidebands spaced by twice the initial spacing ($\sim$50 FSR). At 966 mA another abrupt transition leads to the evolution into the single-FSR-spaced comb regime characterized by prominent harmonic peaks observed altogether with interleaving Fabry-Perot modes. A narrow intermodal beatnote is observed at the cavity roundtrip frequency in this regime (Figure~2b). The main harmonic peaks are separated again by $\sim$25 FSR. Further increase in pump current leads to the transition into the high-phase-noise regime at 1260 mA characterized by laser modes populating all adjacent Fabry-Perot resonances and an intermodal beatnote acquiring a noise pedestal (Figure~2b). The spectral evolution of QCL$_1$ was reported in Figure 6a of Ref.~\citenum{Mansuripur2016}.

\section{Multiheterodyne detection experiments}
\subsection{Frequency counting}
\label{sec_injectionmode}
A key feature of the downconverted RF spectrum produced in a multiheterodyne beating experiment (see Figure~3e) is that its spacing is directly linked to the spacing of the two optical spectra producing it. Specifically, given the repetition rates $f_{\mathrm{rep}, \, 1}$ and $f_{\mathrm{rep}, \, 2}$ of the harmonic state of QCL$_1$ and single-FSR-spaced comb state of QCL$_2$, respectively, the spacing of the multiheterodyne spectrum is $f_{\mathrm{rep}, \, 1}\, \mathrm{mod} \,f_{\mathrm{rep}, \, 2}$. Therefore, measuring the spacing uniformity of the multiheterodyne spectrum allows one to directly verify the equidistant spacing of the harmonic state. 
Three selected beatnotes of the downconverted RF spectrum generated in the multiherodyne experiment described in the main text are shown in Figure S~\ref{figtraces}a. Their frequencies, $f_1$, $f_2$ and $f_3$, lie within the passbands of the filters and are measured with three synchronized $\Lambda$-type frequency counters (gate time = 10 ms, duty cycle = 50\%) during a time interval of 8 s. As can be observed in Figure S~\ref{figtraces}b-d the temporal traces of $f_1$, $f_2$ and $f_3$ exhibit a high degree of correlation despite sizeable, low frequency oscillations of the order of few MHz caused by thermal fluctuations occurring in the lasers. For every set of data points in the traces a value of deviation from equidistant spacing can be defined as $\varepsilon(t)=f_3(t)-2f_2(t)+f_1(t)$. The statistical analysis of the oscillations of $\varepsilon(t)$ allows one to construct a histogram to verify the spacing uniformity of the RF comb. We derive the fractional frequency stability of the dual-comb system using the Allan deviation algorithm for different averaging time intervals and normalized to the optical bandwidth of the measurement (801 GHz). To be precise, the Allan deviation is defined for frequency measurements acquired using zero-dead-time counters~\cite{Allan1997}. Measurements involving $\Lambda$-type counters with non-zero dead time, as in the present study, have reduced sensitivity to fluctuations at high Fourier frequencies (as compared to the reciprocal of the gate time), possibly resulting in an incorrect estimate of the Allan deviation~\cite{Dawkins2007}. However, in terms of the present work the fractional frequency stability is suitable to identify the nature of the low-frequency noise of the system and to compare the relative stability of the downconverted comb among different detection schemes.

\subsection{External detection mode}
\label{sec_externalbeating}
To complement the multiheterodyne experiment in self-detection mode discussed in the main text we carried out an experiment in external detection mode, where the beating of the optical fields emitted from the two QCLs occurs on an external fast photodetector ensuring the uncoupling of the lasers. The optical set-up used in this experiment is shown in Figure S~\ref{figextdetection}a, while the RF circuit is the same of Figure~3f. Qualitatively, the same results of the experiment carried out in injection mode are obtained in the arrangement of external beating (Figure S~\ref{figextdetection}b,c) confirming the spacing uniformity of the harmonic state, however with a somewhat smaller relative accuracy of $5\times10^{-11}$ due to the larger instability of the multiheterodyne signal occurring in this scheme, which causes more frequent fluctuations of the beatnotes outside the passband of the filters and, consequently, a larger number of missed counts by the frequency counters.

\section{Theoretical model of harmonic comb formation}
\label{sec_theory}
In this section we outline the simplest model of the nonlinear mode coupling in a laser with a two-level active medium leading to the formation of the harmonic comb spectrum featuring a high degree of parametric suppression of intermediate cavity modes. For the sake of simplicity we consider the case of a high-Q symmetric Fabry-Perot cavity where an orthogonal set of cavity modes can be defined, as opposed to a leaky cavity with HR/AR-coated facets as in the devices used in our experimental study. We will calculate the linear parametric gain of the sidebands in the presence of a strong central laser mode, and assume that the actual sidebands in the nonlinear regime will develop at the maximum of the gain spectrum. We note the difference of this approach with respect to that of other reported models investigating the impact of FWM on mode-coupling and modal phase relationship in single-FSR-spaced QCL combs, which also rely on Maxwell-Bloch equations but consider an electric field constituted by a large number of optical modes spaced by the cavity roundtrip frequency~\cite{Khurgin2014, Villares2015theory}. The space-time-domain simulations of a harmonic comb formation with a more realistic model for the QCL active region and arbitrary cavity geometry are underway and will be reported elsewhere.   

We develop a linear theory of the sideband formation due to FWM within a 1D cavity model described by the Maxwell-Bloch equations
\begin{align}
\label{Eq:MB}
\partial_t \rho_{ul} = - \left( i \omega_{ul} + 
\frac{1}{T_2} \right) \rho_{ul} - i \frac{d E}{\hbar} \Delta ,  \nonumber \\
\partial_t \Delta = - \frac{\Delta-\Delta_p}{T_1} - 2 i \frac{d E}{\hbar} (\rho_{ul} - \rho_{ul}^\ast) + D \frac{\partial^2\Delta}{\partial z^2} ,  \nonumber \\
\partial_z^2 E - \frac{n^2}{c^2} \partial_t^2 E = \kappa d \partial_t^2 (\rho_{ul} + \rho_{ul}^\ast) ,
\end{align}
where $D$ is the diffusion coefficient of the population inversion $\Delta$, $E$ is the electric field, $\rho_{ul}$ is the off-diagonal density matrix element at the laser transition, $n$ is the effective refractive index of a given transverse waveguide mode, $\kappa = 4\pi \Gamma /  c^2 L_p$ with $\Gamma$ being the optical confinement factor and  $L_p$ the length of one period of the active region. 

As both facets have the same reflectivity, any single-frequency component is a standing wave. Assuming that the reflection on the facets gives a zero phase shift and the laser cavity is between $z = 0$ and $z = L$,  the spatial structure is $\cos(k z)$, with $k L = \mathrm{integer}\times\pi$. We assume that the laser starts at a central mode and then two weak sidebands appear which have equal detunings  $\delta\omega$ and $-\delta\omega$, as required by energy conservation. Also, the wavevector detunings are equal and opposite between the two sidebands corresponding to the phase matching requirement. If the sidebands are weak, we can keep only the couplings to the first order of the sidebands; then no other frequency components will appear.  So we can use the set of ansatzes  
\begin{align}
\label{Eq:amplitudes}
E &= E_0 \cos(k_0 z) e^{-i\omega_0 t} + E_+ \cos(k_+ z) e^{-i \omega_+ t} + E_- \cos(k_- z) e^{-i \omega_- t}   + \mathrm{c.c.} ,  \nonumber \\
\rho_{ul} &= \eta_0 \cos(k_0 z) e^{-i\omega_0 t} +\eta_+ \cos(k_+ z) e^{-i \omega_+ t} + \eta_- \cos(k_- z) e^{-i \omega_- t} ,  \nonumber \\
\Delta &= \Delta_0 + \Delta_2 \cos(2 k_0 z) \nonumber \\ 
&+ \Delta_+ \cos(\delta k z) e^{-i\delta\omega t} + \Delta_- \cos(\delta k z) e^{i\delta\omega t} \nonumber \\
&+ \left[ \Delta_{2+} \cos((k_0 + k_+) z) e^{-i\delta\omega t} + \Delta_{2-} \cos((k_0 + k_-) z) e^{i\delta\omega t} + \mathrm{c.c.} \right] ,
\end{align}
where $\Delta_- = \Delta_+^\ast$, $k_0 = n(\omega_0) \omega_0/c$, $\omega_\pm = \omega_0 \pm \delta\omega$ and $k_\pm = k_0 \pm \delta k$. Here we have kept only the relevant terms to the first order. We will first solve for the steady state in the presence of the central mode.  Assuming the field of the central mode $E_0$ to be not too strong so that the Rabi frequency $d |E_0|/\hbar$ is smaller than the saturation value $1/\sqrt{T_1 T_2}$, we can keep terms up to the order $ |E_0|^2$. This is equivalent to the $\chi^{(3)}$ approximation for the resonant gain nonlinearity. Proceeding in this way we obtain linear equations for complex sideband amplitudes. 
The phase part of these equations is
\begin{align}
\label{Eq:dispersion}
\left( \frac{n^2(\omega_+)\omega_+^2}{c^2} - k_+^2 \right) = - \kappa d \omega_+^2 \left( \Re[\alpha_{++}] + \Re[\alpha_{+-} e^{i\phi}] \frac{|E_-|}{|E_+|} \right) ,  \nonumber \\
\left( \frac{n^2(\omega_-)\omega_-^2}{c^2} - k_-^2 \right) = - \kappa d \omega_-^2 \left( \Re[\alpha_{--}] + \Re[\alpha_{-+} e^{i\phi}] \frac{|E_+|}{|E_-|} \right) .
\end{align} 
The amplitude equations are obtained by taking into account that in the linear theory both sidebands grow in time exponentially with gain $g$, i.e.~as $e^{gt}$:
\begin{align}
\label{Eq:gain}
2 \frac{ n^2(\omega_+)}{c^2} g = -\kappa d \omega_+ \left( \Im[\alpha_{++}] + \Im[\alpha_{+-} e^{i\phi}] \frac{|E_-|}{|E_+|} \right)  , \nonumber \\
2 \frac{ n^2(\omega_-)}{c^2} g = -\kappa d \omega_- \left( \Im[\alpha_{--}] + \Im[\alpha_{-+} e^{i\phi}] \frac{|E_+|}{|E_-|} \right) .
\end{align}
The parameters $\alpha$ are related to the susceptibility tensor as 
\begin{align}
&\eta_+ = \alpha_{++} E_+ + \alpha_{+-} E_-^\ast ,  \nonumber \\
&\eta_- = \alpha_{--} E_- + \alpha_{-+} E_+^\ast ,
\end{align}
and the expressions of $\alpha$ are
\begin{align}
&\alpha_{++} = \frac{d}{\hbar} \frac{\Delta_{th}}{\delta\omega+i/T_2} \left[ 1 + \left(\frac{d|E_0|}{\hbar}\right)^2 \left( T_g T_2 + \frac{1}{2}  \left( \frac{1}{\delta\omega+i/T_1} + \frac{1}{\delta\omega+i/T_g} \right) \left( \frac{1}{\delta\omega+i/T_2} - i T_2 \right) \right) \right] , \nonumber \\
&\alpha_{+-} = \frac{1}{2} \frac{d}{\hbar} \frac{\Delta_{th}}{\delta\omega+i/T_2} \left(\frac{d E_0}{\hbar}\right)^2 \frac{1}{\delta\omega+i/T_1} \left( -i T_2 + \frac{1}{\delta\omega+i/T_2} \right) , \nonumber \\
&\alpha_{--} = \alpha_{++}|_{\delta\omega \rightarrow -\delta\omega} , \nonumber \\
&\alpha_{-+} = \alpha_{+-}|_{\delta\omega \rightarrow -\delta\omega}  ,
\end{align}
where ${\Delta_{th}}$ is the threshold population inversion, and $T_g^{-1} = T_1^{-1} + 4 k_0^2 D$.

Eqs.~(\ref{Eq:dispersion}) and (\ref{Eq:gain}) contain 5 unknowns: relative phase $\phi = \arg(E_-^\ast) -\arg(E_+)$,  $\delta\omega$, $\delta k$, $|E_-|/|E_+|$ and $g$. One can solve for any 4 parameters as a function of the fifth. In the main text we plot the gain spectrum as a function of $\delta\omega$ considering a pumping 10\% larger than the lasing threshold (Fig.~4).
In these plots we include the GVD effect due to the material and waveguide through the frequency dependence of the refractive index $n(\omega) = 3.23 + 0.5 c \cdot \beta_{GVD}  \cdot (\omega - \omega_0)$ and use all other parameters for this laser~\cite{Mansuripur2016}: emission wavelength $\lambda = 4.5$ $\mu$m, gain recovery time $T_1 = 1.7$ ps, dephasing time $T_2 = 74$ fs, dipole moment $d = 1.63$ nm, and sub-threshold GVD $\beta_{GVD} = 400$ fs$^2$/mm.  The gain-related GVD and higher-order dispersive terms are included automatically in the model. 

In our analysis, the two sidebands have the same frequency detuning. However, the GVD disperses the modes producing mismatched frequency detunings. This mismatch is pulled back by the interaction between the sidebands and the central mode. We can define a frequency pulling, which is equal to the frequency mismatch due to GVD. It is derived to be
\begin{equation}
\Delta\omega_{pull}\left(\beta_{GVD}\right) = \frac{c \delta\omega^2 \beta_{GVD}}{n(\omega_0) + \frac{1}{2} c \omega_0 \beta_{GVD}} .
\end{equation}
The frequency pulling is plotted as a function of subthreshold GVD in Fig.~4c of the main text.

The plot of the gain spectrum shows a large separation of the gain peaks: around 200 GHz from the central mode (26 FSR of a 6 mm long cavity), in qualitative agreement with the experiments. With increasing magnitude of the material/waveguide GVD, the solution with sideband gain maxima satisfying energy and phase matching can be found up to the value $\beta_{GVD} \sim 6000$ fs$^2$/mm. Above this value of GVD, no such solution exists, indicating that FWM cannot overcome the dispersion and cannot support harmonic modelocking.

\newpage

\section*{Figures}

\begin{figure}[H]
    \centering
    \includegraphics[width=1\columnwidth]{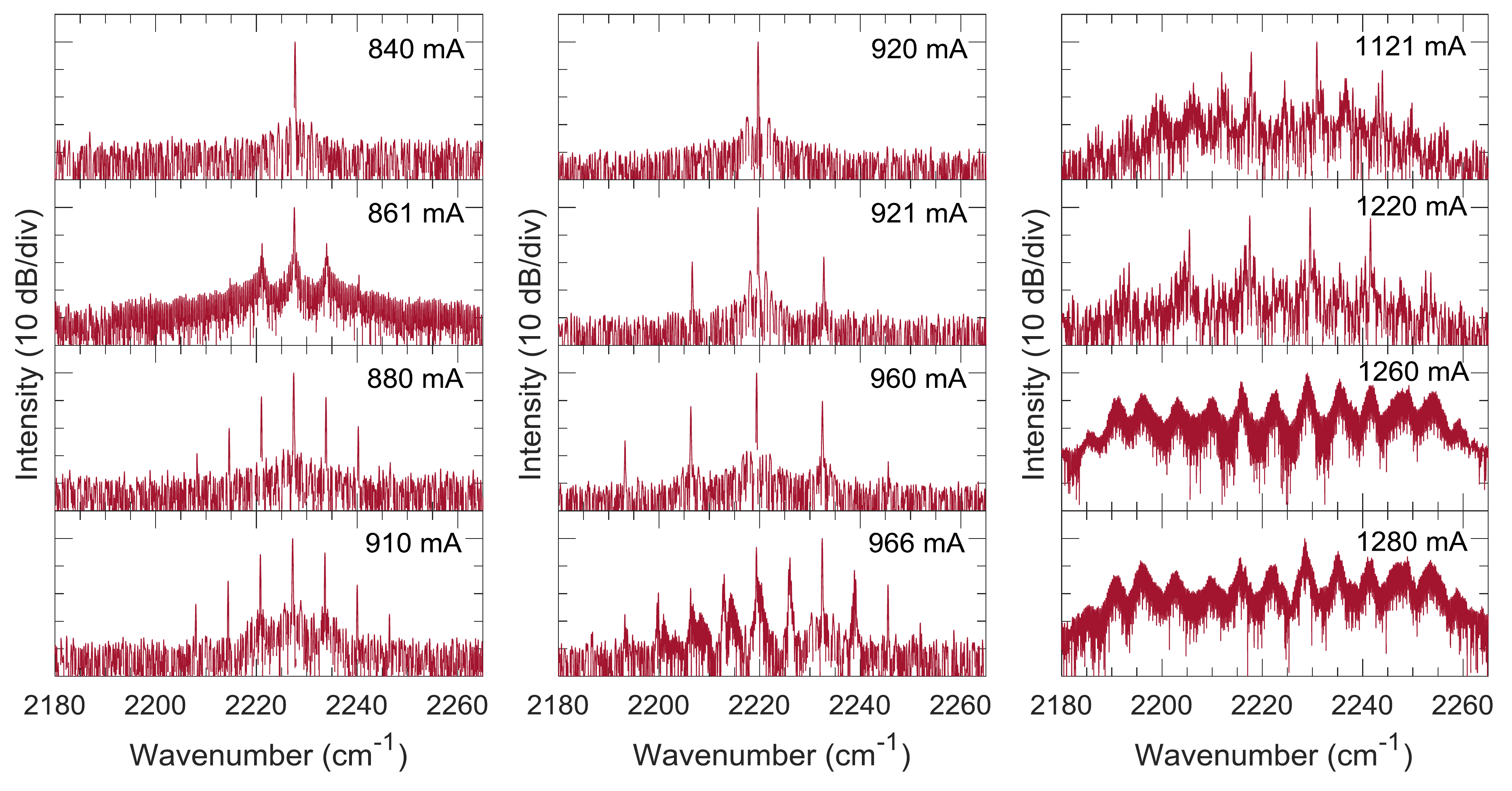}
    \caption{Spectral evolution of a QCL as a function of the injection in the device starting from the lasing threshold and increasing the current.}
	\label{figevolutionall}
\end{figure}

\newpage

\begin{figure}[H]
    \centering
    \includegraphics[width=0.5\columnwidth]{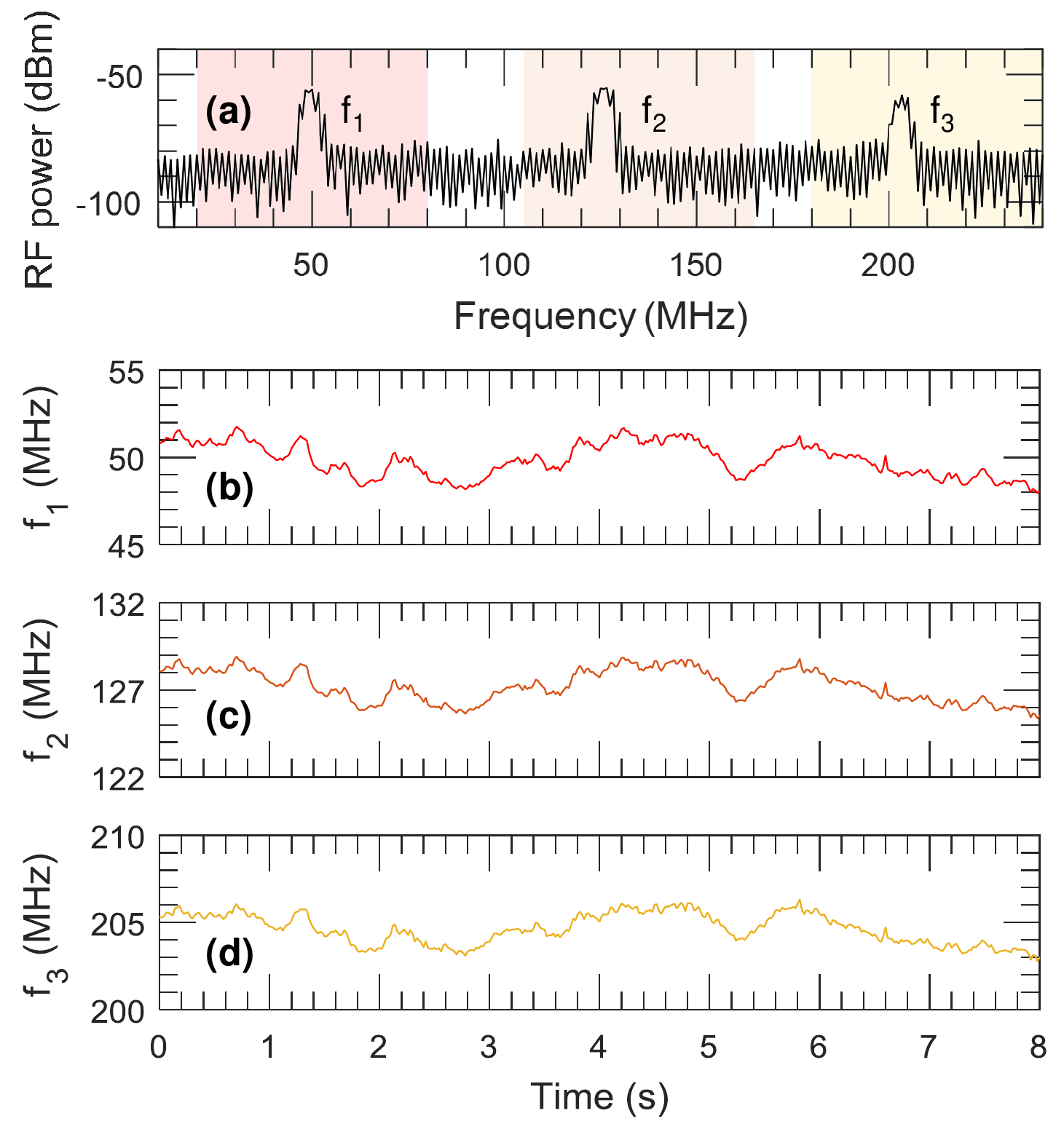}
    \caption{\textbf{Frequency counting}. (a) Three selected beatnotes of the RF spectrum generated in the multiheterodyne experiment shown in Figure~3. The colored rectangles correspond to the different bandwidths of the RF filters. (b)-(d) Frequency traces measured by the synchronized frequency counters over a time interval of 8 s (gate time: 10 ms).}
	\label{figtraces}
\end{figure}

\newpage

\begin{figure*}
    \centering
    \includegraphics[width=1\textwidth]{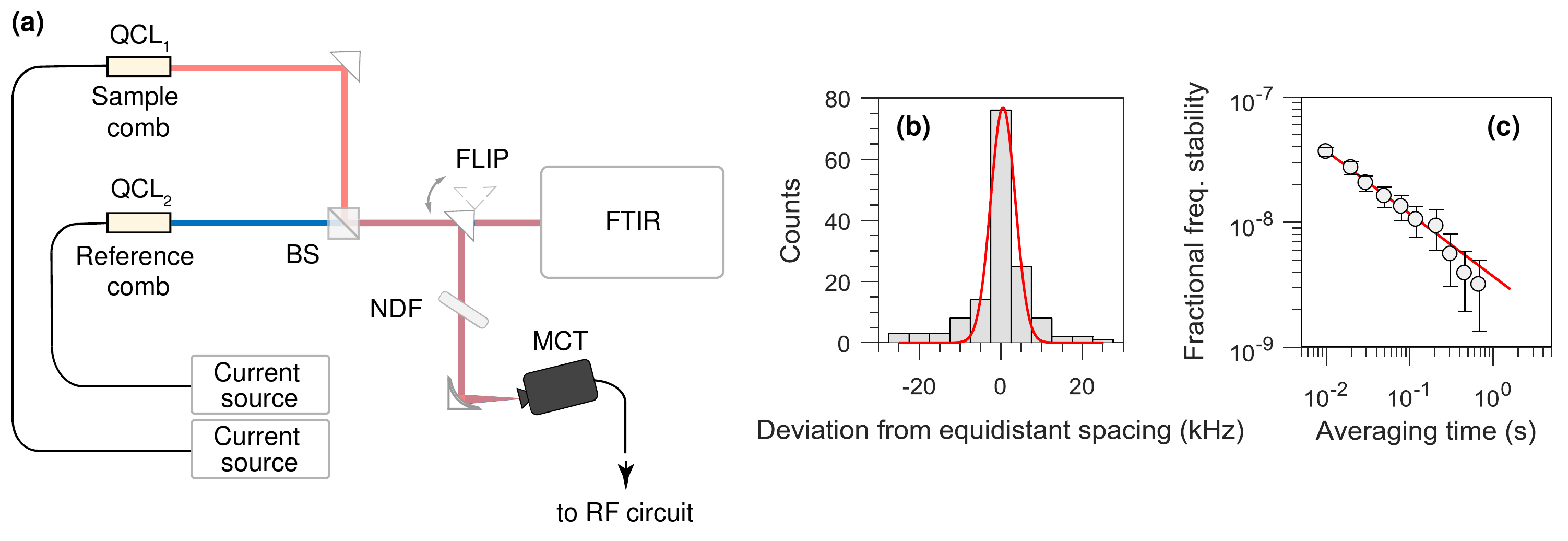}
    \caption{\textbf{Multiheterodyne experiment in external detection mode}. (a) Optical set-up for the assessment of the comb spacing uniformity of the harmonic state in external detection mode. FM, flip mirror; NDF, neutral density filter; FLIP, flip mirror; MCT, HgCdTe detector. (b) Histogram showing the deviation from equidistant spacing of the harmonic state measured with the set-up shown in (a) for a gate time of 10 ms and 147 counts. The parameters of the Gaussian fit (red curve) are $\mu=0.5$ kHz and $\sigma=3.0$ kHz. (c) Fractional frequency stability of the dual-comb system in external detection mode. The power fit (red line) indicates an inverse square root dependence on the averaging time (fitted exponent: $-0.50$).}
	\label{figextdetection}
\end{figure*}
 
\bibliography{harmcombbib}